\documentclass[a4paper]{jpconf}
\usepackage{graphicx}
\begin{document}
\title{ANAIS-112 status: two years results on annual modulation}

\author{J.~Amar{\'e}$^{1,2 }$, S.~Cebri{\'an}$^{1,2 }$, D. Cintas$^{1,2 }$,~I.~Coarasa$^{1,2 }$, E.~Garc\'{\i}a$^{1,2 }$, M.~Mart\'{\i}nez$^{1,2,3 }$, M.A. Oliv{\'a}n$^{1,2,4 }$, Y.~Ortigoza$^{1,2 }$, A.~Ortiz~de~Sol{\'o}rzano$^{1,2 }$, J.~Puimed{\'o}n$^{1,2 }$, A.~Salinas$^{1,2 }$, M.L.~Sarsa$^{1,2 }$ and P.~Villar$^{1,2 }$}

\address{$^{1 }$Centro de Astropart{\'{\i}}culas y F{\'{\i}}sica de Altas Energ{\'{\i}}as (CAPA), Universidad de Zaragoza, Pedro Cerbuna 12, 50009 Zaragoza, Spain}
\address{$^{2 }$Laboratorio Subterr{\'a}neo de Canfranc, Paseo de los Ayerbe s.n., 22880 Canfranc Estaci{\'o}n, Huesca, Spain}
\address{$^{3 }$Fundaci{\'o}n ARAID, Av. de Ranillas 1D,  50018 Zaragoza, Spain}
\address{$^{4 }$Fundaci{\'o}n CIRCE, 50018, Zaragoza, Spain}
\ead{mlsarsa@unizar.es}

\begin{abstract}
ANAIS (Annual modulation with NaI Scintillators) is a dark matter direct detection experiment located at the Canfranc Underground Laboratory (LSC), in Spain. The goal is to confirm or refute in a model independent way the DAMA/LIBRA positive result: an annual modulation in the low-energy detection rate compatible with the expected signal induced by
dark matter particles in the galactic halo. This signal, observed for about 20 years, is in strong
tension with the negative results of other very sensitive experiments, but a direct comparison
using the same target material, NaI(Tl), was still lacking. ANAIS-112, consisting of 112.5 kg
of NaI(Tl) scintillators, is taking data at the LSC since August 2017. Here we present the preliminary annual modulation analysis corresponding to two years of data (exposure of 220.69 kg$\times$y) and the ANAIS-112
projected sensitivity for the scheduled 5 y of operation.

\end{abstract}

\section{Introduction}
Large experimental efforts devoted to unraveling the nature of the dark matter (DM) particles have been carried out, either by direct~\cite{Liu}, indirect~\cite{Conrad} or accelerator searches~\cite{Buchmueller}, which are complementary to each other. For about twenty years, the DAMA/LIBRA collaboration has been claiming the observation of an annual modulation in the detection rate, which fulfills all of the requirements expected for the contribution of weakly interacting DM particles, distributed in the Milky Way halo. DAMA/LIBRA detector is installed at Gran Sasso Underground Laboratory (LNGS), in Italy, and it consists of highly radiopure NaI(Tl) scintillators, having a total mass of 250~kg~\cite{DAMA}. The current statistical significance of the DAMA/LIBRA modulation result reaches the 12$\sigma$ level. However, it has neither been reproduced by any other experiment, nor ruled out in a model independent way~\cite{XENON,LUX,PICO,SUPERCDMS,CRESST,COSINE_Nature}. Compatibility among the different experimental results in most conventional WIMP-DM scenarios is actually disfavored~\cite{Baum,Kang}.

Other experiments using the same target are crucial to ascertain whether the DAMA/LIBRA positive signal is a signature of the halo dark matter particles or some systematics.  There are several efforts around the world pursuing this goal~\cite{SABRE,COSINUS,PICOLON}. COSINE-100 and ANAIS-112 experiments are presently in data taking phase, and both have already published preliminary analysis. COSINE-100 experiment is located at the Yangyang Underground Laboratory, in South Korea and it consists of 106~kg of NaI(Tl) detectors, immersed in a liquid scintillator, which allows the identification and subsequent reduction of some radioactive backgrounds. An analysis of the first 59.5 live days using about 61~kg effective mass allowed the COSINE collaboration to exclude the 3$\sigma$ DAMA/LIBRA region at 90$\%$ C.L. in a model dependent way~\cite{COSINE_Nature}, and the annual modulation analysis of 1.7~years, with a total exposure of 97.7 kg$\times$y, recently published is compatible at 1$\sigma$ both with DAMA/LIBRA signal and with the absence of modulation~\cite{COSINE_PRL}.

ANAIS-112 experiment is taking data at the Canfranc Underground Laboratory in Spain since August 2017. It consists of 112.5~kg of NaI(Tl)detectors, disposed in a 3x3 array of modules, 12.5~kg each. Most relevant features of ANAIS modules, built by Alpha Spectra Inc., are the Mylar window built-in to allow low energy calibration with external sources, and outstanding optical quality, which added to the high-efficiency Hamamatsu photomultipliers (PMTs) enable a light collection at the level of 15~photoelectrons (phe) per keV in all the modules. The ANAIS-112 shielding consists of 10~cm of archaeological lead, 20~cm of low activity lead, an anti-radon box (kept under overpressure with radon-free nitrogen gas) and 40~cm of a combination of water tanks and polyethylene bricks. An active muon veto made up of 16 plastic scintillators covers the top and sides of the set-up~\cite{ANAIS2019_perf}. ANAIS-112 DAQ hardware and software is robust and it has been fine-tuned in the operation of several prototypes. The signals from the two PMTs coupled to each NaI(Tl)-module are fully processed: signal is divided into a trigger signal, a high-energy signal, and for the low-energy signal the waveform is digitized at 2GS/s with high resolution. Trigger is done by the coincidence (logical AND) of the two PMT trigger signals in a 200~ns window, while the trigger of each PMT is at phe level. $^{109}$Cd sources are used every two weeks to calibrate the experiment and correct, if necessary, for possible gain drifts. Energy calibration is done by combining the information from the $^{109}$Cd lines and $^{40}$K and $^{22}$Na crystal contaminations energy depositions at very low energy, being the latter tagged by coincidences with high-energy gammas, and corresponding to 3.2 and 0.87 keV. This procedure allows calibrating ANAIS-112 down to the threshold with high accuracy. Triggering below 1~keV\footnote{All the energies used throughout this paper are electron equivalent energies.} with almost 100$\%$ efficiency is guaranteed by the observation of the $^{22}$Na events population. However, analysis threshold is set up at 1~keV because of the PMT-related events rejection, which forces to establish events selection criteria whose acceptance efficiencies for bulk scintillation events decrease down to about 15$\%$ at 1~keV~\cite{ANAIS2019_perf}.

\section{Annual modulation analysis and results} 
Energy and time distribution from single-hit events in the ROI is kept blinded in our analysis protocol. After the first year of data taking we unblinded 10$\%$ of those events, chosen randomly distributed days along the data taking, for fine tuning the events rejection procedures and carry-out general background assessment and sensitivity estimates. All of this information was published~\cite{ANAIS2019_perf,ANAIS2019_bkg,ANAIS2019_sens}. After 1.5~years, using the designed analysis protocol, the unblinding of the ROI and annual modulation analysis allowed the confirmation of the sensitivity projections, while producing some tension with DAMA/LIBRA annual modulation result~\cite{ANAIS2019_results}. This result corresponded to an effective exposure of 157.55~kg$\times$y.

After 2~years, using the same analysis protocol, we have carried out a new unblinding of ANAIS-112 data and the corresponding annual modulation analysis. While preparing a more complete analysis, we present here the result of a model independent analysis searching for modulation in the same regions as DAMA/LIBRA has published ([1-6]~keV and [2,6]~keV) using an exposure of 220.69 kg$\times$y. Data from all the modules are added together and we use a least-squared fit, modeling the data as:

\begin{equation}
\label{eq:fit}
R(t)=R_0+R_1 \cdot exp(-t/\tau)+S_m \cdot cos(\omega \cdot (t-t_0))
\end{equation}

\begin{figure}
\begin{center}
\includegraphics[width=30pc]{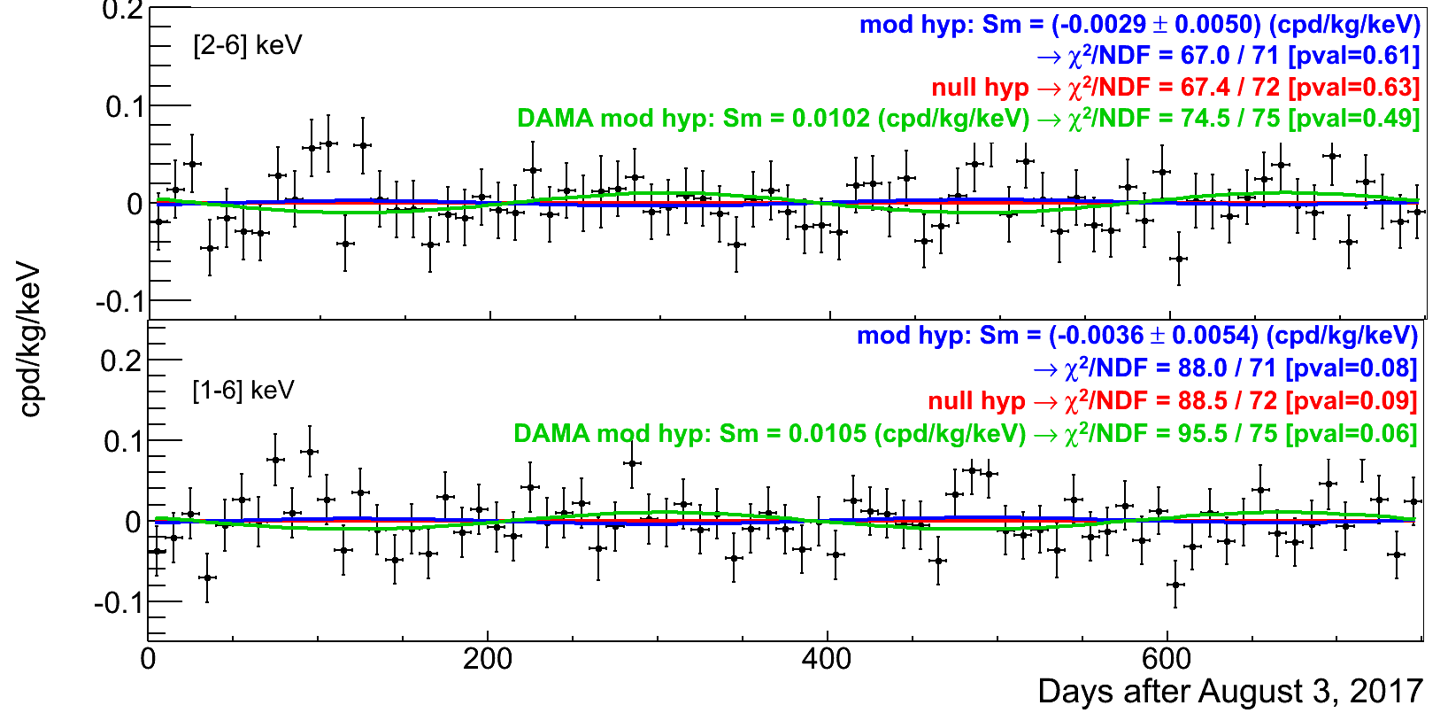}
\end{center}
\caption{\label{fig:residuals}ANAIS-112 fit results for two years of data in energy regions [1-6] and [2-6] keV in the modulation (blue) and null hypothesis (red). Data are displayed after subtracting the constant and exponential functions fitted to eq.~\ref{eq:fit}. DAMA/LIBRA results are shown in green for comparison.}
\vspace{-0.5cm}
\end{figure}

\begin{figure}[h]
\includegraphics[width=18pc]{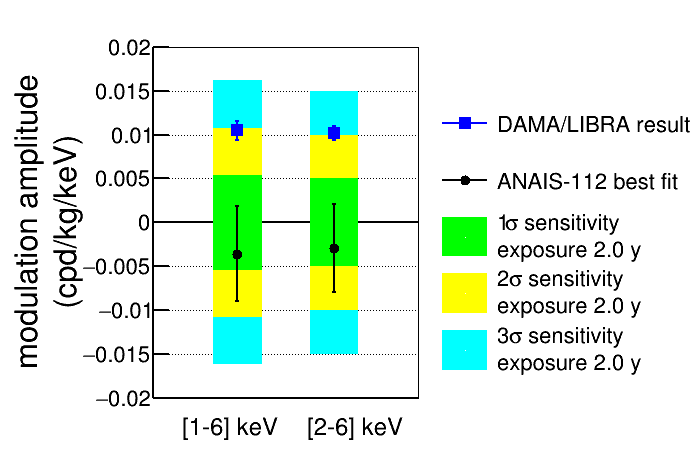}\hspace{2pc}%
\begin{minipage}[b]{14pc}\caption{\label{fig:results}Comparison between ANAIS-112 results on annual modulation using two years of data and DAMA/LIBRA modulation best fit. Estimated sensitivity is shown at different confidence levels as coloured bands: green at 1$\sigma$, yellow at 2$\sigma$, and cyan at 3$\sigma$. }
\end{minipage}
\end{figure}

\begin{figure}[h]
\includegraphics[width=16pc]{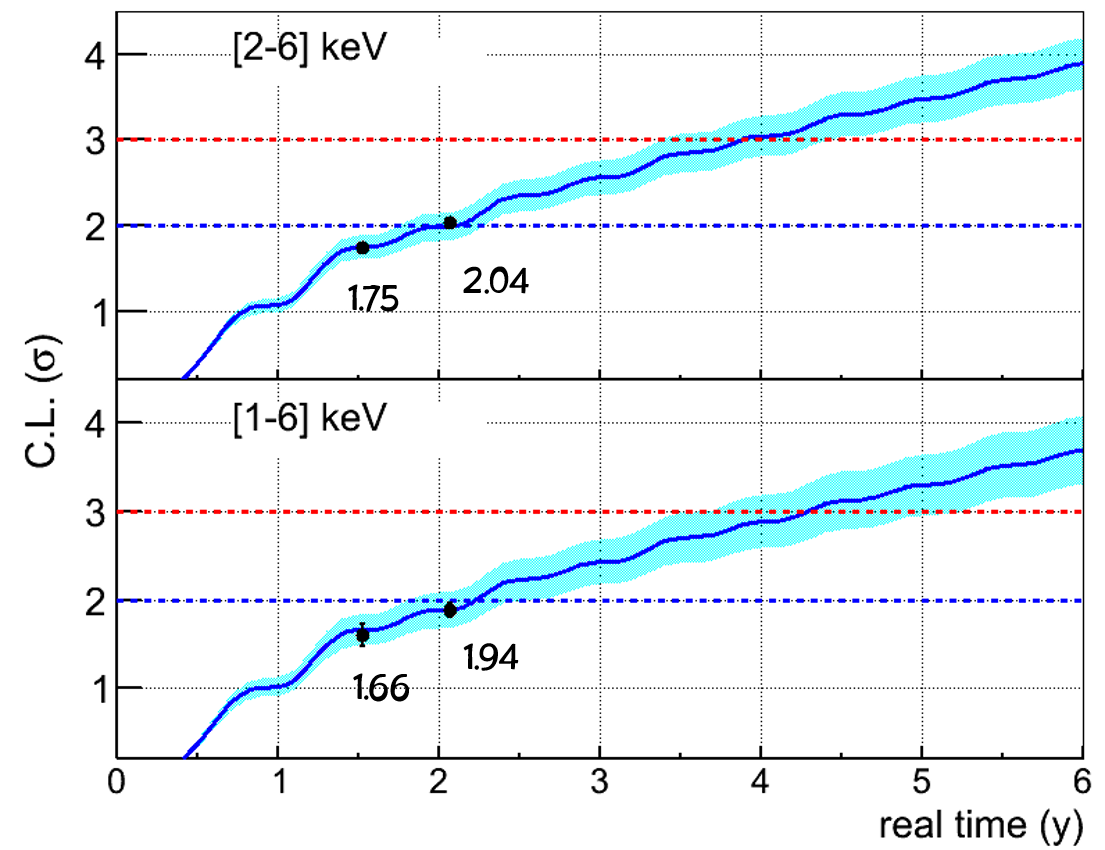}\hspace{2pc}%
\begin{minipage}[b]{14pc}\caption{\label{fig:sens}ANAIS-112 sensitivity to DAMA/LIBRA signal in units of $\sigma$ C.L. as a function of real time. Cyan bands represent the 68$\%~$ C.L. DAMA/LIBRA uncertainty in the modulation amplitude. Black dots and numbers are the experimental sensitivities obtained in the two ANAIS-112 analysis, which correspond to 1.5 and 2 years. }
\end{minipage}
\vspace{-0.4cm} 
\end{figure}

$R_0$ and $R_1$ are free parameters, but $\tau$ is fixed to the value obtained for the time evolution of our background model in the corresponding energy range. Period and phase are fixed to the expected values for the standard galactic halo: 1~y and maximum at June, 2, respectively. $S_m$ is fixed to zero for the null hypothesis and left unconstrained for the modulation hypothesis. Figure~\ref{fig:residuals} shows the fit results of detection rate in the ROI to eq~\ref{eq:fit}. The non-modulated components have been subtracted in the graphical representation. Results are consistent with the null hypothesis, and best fits for the modulation hypothesis are compatible with the absence of modulation within 1$\sigma$. Moreover, those best fits are incompatible with DAMA/LIBRA results at about 2.6$\sigma$. These results, together with the sensitivity projections, are summarized in Figure~\ref{fig:results}. We quote our sensitivity to DAMA/LIBRA as the ratio of DAMA modulation result over the standard deviation on the modulation amplitude derived from ANAIS-112 data ($S_m^{DAMA}/\sigma(S_m)$). Results from two-year data of ANAIS-112 confirm our sensitivity prospects~\cite{ANAIS2019_sens}, being at present at 2$\sigma$ level, and supporting our goal of reaching 3$\sigma$ in 5 years of data taking (see Figure~\ref{fig:sens}).

\section{Conclusions}
ANAIS-112 data taking is progressing smoothly since August 2017, accumulating 220.69~kg$\times$y until beginning of September 2019. The achieved sensitivity to test DAMA/LIBRA result is at 2$\sigma$ level. Best fits ($S_m=-0.0029 \pm 0.0050$ cpd/kg/keV and $-0.0036 \pm 0.0054$ cpd/kg/keV in [2-6] and [1-6] keV energy regions, respectively) are compatible with the absence of modulation, and incompatible with DAMA/LIBRA result at 2.6$\sigma$. The confirmation of our sensitivity prospects implied by this result, guarantees our ability to test DAMA/LIBRA at 3$\sigma$ in less than three years from now, bringing a new light into this long standing puzzle.  

\section*{Acknowledgments}

This work has been financially supported by the Spanish Ministerio de Econom{\'\i}a y Competitividad and the European Regional Development Fund (MINECO-FEDER) under grant FPA2017-83133-P, the Consolider-Ingenio 2010 Programme under grants MultiDark CSD2009-00064 and CPAN CSD2007-00042, the LSC Consortium, and the Gobierno de Arag{\'o}n and the European Social Fund (Group in Nuclear and Astroparticle Physics and I.~Coarasa predoctoral grant). We thank the support of the Spanish Red Consolider MultiDark FPA2017-90566-REDC. Authors would like to acknowledge the use of Servicio General de Apoyo a la Investigaci\'on-SAI, Universidad de Zaragoza and technical support from LSC and GIFNA staff.

\end{document}